\documentclass[runningheads]{llncs}
\usepackage{xurl}
\usepackage{cite}
\usepackage{amsmath}
\usepackage{hyperref}
\usepackage{graphicx}
\usepackage{subcaption}
\usepackage{multirow}
\usepackage{booktabs}
\usepackage{xcolor}
\usepackage{wrapfig}

\newcommand{\DOI}[1]{\textbf{DOI}: #1}
\begin{document}
\title{Visualizing Environmental Justice Issues in Urban Areas with a Community-based Approach}
\titlerunning{SDSS 2021}
% \titlerunning{Visualizing environmental justice issues in urban areas}
\author{Joel Flax-Hatch\inst{1} \and
Sanjana Srabanti\inst{2} \and
Fabio Miranda\inst{2} \and\\
Apostolis Sambanis\inst{3} \and
Michael D. Cailas \inst{1}}
\authorrunning{J. Flax-Hatch et al.}
\institute{Environmental and Occupational Health Sciences, School of Public Health, University of Illinois at Chicago\\\email{\{jflaxh2, mihalis\}@uic.edu} \and
Department of Computer Science, College of Engineering,\\University of Illinois at Chicago\\
\email{\{ssraba2, fabiom\}@uic.edu}\and
Health Policy and Administration, School of Public Health,\\University of Illinois at Chicago\\
\email{asamba2@uic.edu}}
\maketitle              % typeset the header of the contribution
\begin{abstract}
According to environmental justice, environmental degradation and benefits should not be disproportionately shared between communities. 
Identifying disparities in the spatial distribution of environmental degradation is therefore a prerequisite for validating the state of environmental justice in a geographic region.
Under ideal circumstances, environmental risk assessment is a preferred metric, but only when exposure levels have been quantified reliably after estimating the risk.
In this study, we adopt a proximity burden metric caused by adjacent hazardous sources, allowing us to evaluate the environmental burden distribution and vulnerability to pollution sources.
In close collaboration with a predominantly Latinx community in Chicago, we highlight the usefulness of our approach through a case study that shows how certain community areas in the city are likely to bear a disproportionate burden of environmental pollution caused by industrial roads.

% The abstract should briefly summarize the contents of the paper in 15--250 words.

\end{abstract}
\DOI{\url{https://doi.org/10.25436/E2Z30J}}
\vspace{-0.5cm}
\section{Introduction}
\vspace{-0.3cm}
Environmental justice (EJ) concepts can be traced back to the American Civil Rights movement of the 1960s. These concepts gained major impetus in the 1990s with the declaration of the 17 principles during the First National People of Color Environmental Leadership Summit held in Washington (1991) and the Executive Order 12898 (Federal Actions to Address Environmental Justice in Minority Populations and Low-Income Populations; 1994). One fundamental principle is that communities must not disproportionately share environmental degradation or benefits. In this context, environmental degradation implies the existence of stationary or mobile hazardous sources, an exposure pathway, and a population bearing the potential health impact(s) caused by these sources.
According to the Environmental Protection Agency~\cite{epajustice}, environmental justice will be achieved ``\emph{when everyone enjoys the same degree of protection from environmental and health hazards}'' and, on top of that, ``\emph{equal access to the decision-making process to have a healthy environment in which to live, learn, and work.}''

\begin{wrapfigure}{r}{0.25\textwidth}
    \vspace{-1.0cm}
    \centering
    \includegraphics[width=0.9\linewidth]{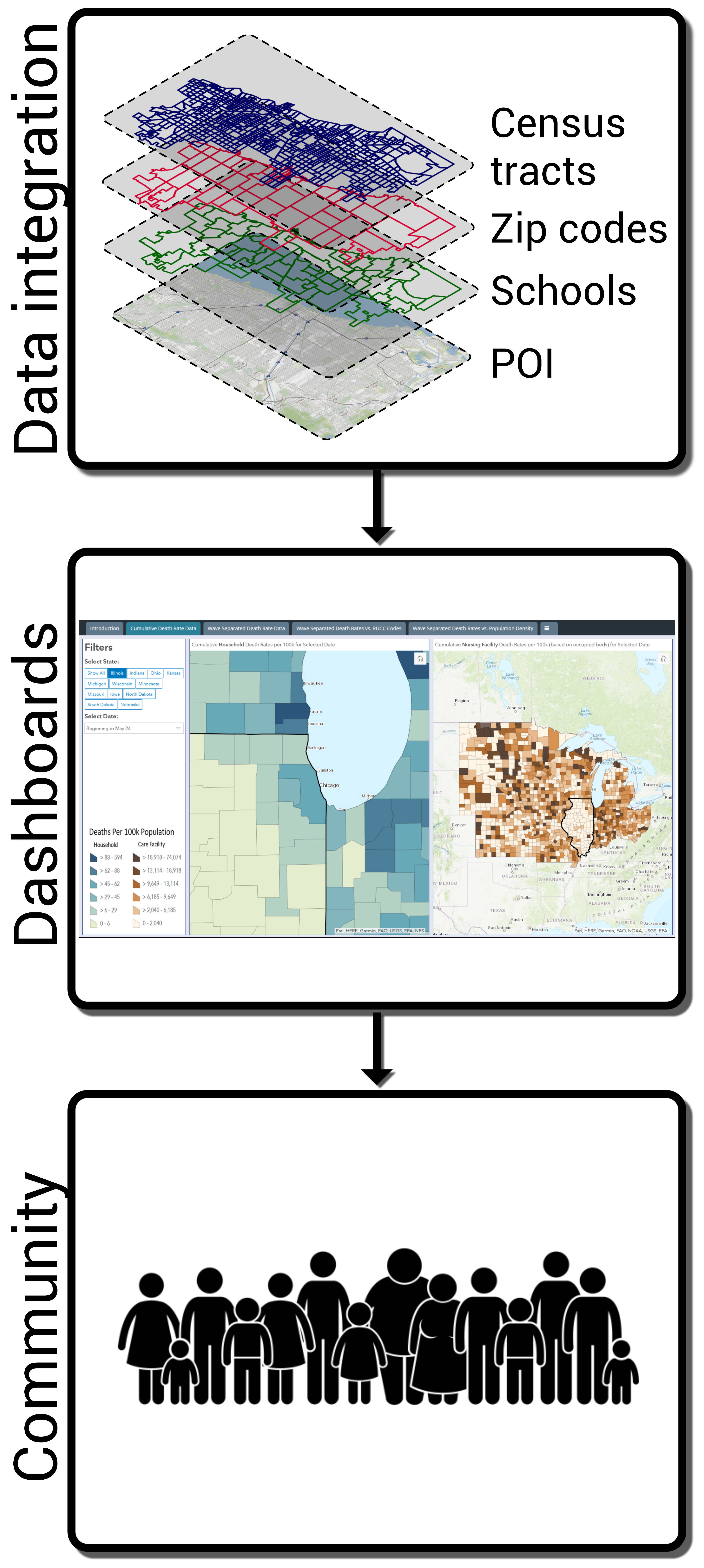}
    \caption{Community-oriented approach.}
    \label{fig:overview}
    \vspace{-0.8cm}
\end{wrapfigure}

A prerequisite for validating the state of EJ in a geographic region is identifying disparities in the spatial distribution of environmental degradation. This implies that a metric must be adopted to represent the level of degradation. Under ideal conditions, a risk estimate is a preferable metric; however, this requires a reliable quantification of the exposure level after conducting an environmental risk assessment~\cite{us2015human}. 
This task poses insurmountable difficulties in EJ validation studies since communities are exposed to many stationary and mobile hazardous sources, and it is difficult to quantify the exposures stemming from each one of them reliably. To overcome this limitation, the proximity burden metric caused by adjacent hazardous sources is introduced, requiring the coordinates of the target population (e.g., the centroid of a census tract)~\cite{brender2011residential}.
In this study, we consider a series of hazardous sources from toxic release inventory facilities, rail yards and brownfields.

As recently pointed out by Mah~\cite{doi:10.1080/23251042.2016.1220849}, big data can play an important role in understanding toxic exposure landscapes across different temporal and spatial scales. This must be accompanied, however, by the right techniques and approaches that render new voices visible and ensure that patterns of exclusion are not reproduced.
Even though underrepresented communities have witnessed and felt disparities through their lived experience, they often do not have the platform to raise their concerns. It is important then to design new methods to increase community engagement with local authorities.
Our focus in this paper is the development of a proximity metric that uses data and considers an underrepresented community in Chicago.
We use as a target population, kindergarten to 8th-grade school children.
% A major innovation of this study is to use as a target population kindergarten (age 5 to 6) to 8th-grade school children (henceforth K-8).  
With this selection, a geographically well-defined location for the target population becomes available (i.e., school addresses). Moreover, this population is age homogeneous, vulnerable to pollution sources, and relatively stationary for many years. To optimize the residency requirement, only the children in the neighborhood schools were included in this study since most of them are likely to dwell in the local communities where the schools are.

In this paper, we first introduce our collective proximity burden (CPB) metric, and discuss the main sources of data used in the study (Fig.~\ref{fig:overview}).
We briefly discuss the creation of a visualization dashboard created after close interactions with community groups and policy makers.
We highlight the usefulness of our approach through a case study that shows how certain community areas in Chicago are likely to bear a disproportionate burden of environmental pollution caused by industrial roads.

\vspace{-0.3cm}
\section{Methodology}
\vspace{-0.3cm}

\subsection{Collective proximity burden}
\vspace{-0.2cm}
To establish a meaningful proximity metric of the environmental burden distribution, we introduce the school proximity burden score (henceforth, for brevity, proximity burden). For each school $i$, the 1-mile proximity burden score is:

\vspace{-0.25cm}
\begin{equation}
    (Proximity Burden)_i = (PSS \times Hs)_i
\end{equation}
where: $PSS$ is the percentage of neighborhood school students (out of the total student population) in each school $i$, and $Hs$ is the number of hazardous sources near school $i$, within a 1-mile radius.
This metric establishes the distribution of the 1-mile proximity burden for each school, $i$, as a relative (to the other schools) score in the study area (i.e., City of Chicago). 
We chose a 1-mile radius because it represents a typical 15-20 minute walk.

To communicate the environmental burden reality in understandable and geographically identifiable terms, aggregation was performed at a Chicago community area level. The collective proximity burden for all the schools in each community area provides a metric to study the exposure disparities to environmental hazards levied on the most sensitive population of these communities. The collective proximity burden (CPB) at a Chicago community area scale, $z$, is defined as follows:

\vspace{-0.25cm}
\begin{equation}
CPB_z = \sum_{i=1}^{n_z}(Proximity Burden)_i = \sum_{i=1}^{n_z}(PSS \times Hs)_i
\end{equation}
where $n_z$ is the number of schools in the community area $z$.

The CPB establishes a comparison metric for each community area regarding the proximity burden on its schools. For example, New City, a community area in the Southwest section of Chicago, has ten public schools (i.e., $n_z$ = 10). Conceptually, the CPB score provides an estimate of the hazard distribution that each community area bears due to the proximity of its schools to hazardous sources within a 1-mile radius.

The case study presented in the following section depicts the industrial roads burden distribution to public schools in Chicago. This is one of the five hazardous sources studied in a series of Midwest Comprehensive Visualization Dashboards (MCVD) focusing on environmental justice issues in the Chicago region~\cite{MCVDa,MCVDb}. 
The other sources were: the toxic release inventory facilities (TRI), the Risk-Screening Environmental Indicator hazard for each TRI facility, rail yards with six of the eight major rail hubs located in the southwest section of the city, and brownfields.
The primary objective of these dashboards is to create visualizations that lead to operational insights supporting data-driven decisions to resolve the EJ issues in large urban settings. Due to the potential public health implications related to the unequal distribution of these pollution exposure sources, we adopted a community-based participatory design approach to ensure, as a minimum, that the representations of data and findings are understandable to the public. 
The visualization for the case study presented in the following section is from the interactive MCVD~\cite{MCVDb}, which has been formulated based on a long period of interactions with community groups and policymakers. The dashboard was widely used by the community and the high number of views and downloads recorded by the UIC usage metric system as well as the coverage it received from major news outlets (e.g., Chicago Tribune~\cite{ChicagoTribune}) testify to its success in improving public discourse and engaging city agents in this important issue. 

%The wide use of these dashboards by the community is documented by the high number of views and downloads recorded by the UIC usage metric system~\cite{indigo} as well as the coverage it received from major news outlets (e.g., Chicago Tribune~\cite{ChicagoTribune}).

%The high number of views and downloads recorded by the UIC usage metric system~\cite{indigo} as well as the coverage these dashboards received from major news outlets (e.g., Chicago Tribune~\cite{ChicagoTribune}), highlight their wide usage by the community. 

\vspace{-0.5cm}
\subsection{Data sources}
\vspace{-0.3cm}
In this study, we use a number of publicly available data sets to analyze the disparities in exposure to environmental hazards among kindergarteners (ages 5 to 6) to eighth graders of Chicago public schools. The School Profile Information provided by Chicago Public Schools (CPS) is used as the primary source of demographic data about the population under study. The socioeconomic indicators were obtained from the 2018 U.S. Census Bureau American Community Survey. 
The geographic information describing the neighborhood characteristics, including major roads, community areas and industrial corridors were obtained from the Chicago Data Portal.
%
% \begin{itemize}
%     \item Chicago Public Schools (CPS) - School Profile Information.
%     \item The socioeconomic data used to study the characteristics of the study area were obtain from the U.S. Census Bureau American Community Survey; 2018 release of 5-year estimates.
%     \item Shapefiles were specialized using NAD 1983 (2011) State plane Illinois East Fips 1201.
%     \item Major roads shapefile collected from the Chicago Data Portal.
%     \item Industrial Corridors were collected from the Chicago Data Portal.
%     \item Community areas were collected from the Chicago Data Portal.
% \end{itemize}
%
Data preparation and preliminary analysis was performed with the IBM SPSS Modeller 18.2.1.
Geospatial data integration, mapping, and spatial analysis were performed using ESRI’s ArcGIS Pro.

\vspace{-0.5cm}
\section{Case study}
\vspace{-0.3cm}
The hazardous source considered in this case study is the burden estimated from the industrial roads at a 1-mile radius from schools (see Eq. 1). The hazardous source is the total kilometers of heavy traffic roads (i.e., roads classified by the city of Chicago) within the industrial corridors of Chicago.

\begin{wrapfigure}{r}{0.5\textwidth}
    \vspace{-0.5cm}
    \centering
    \includegraphics[width=0.5\textwidth]{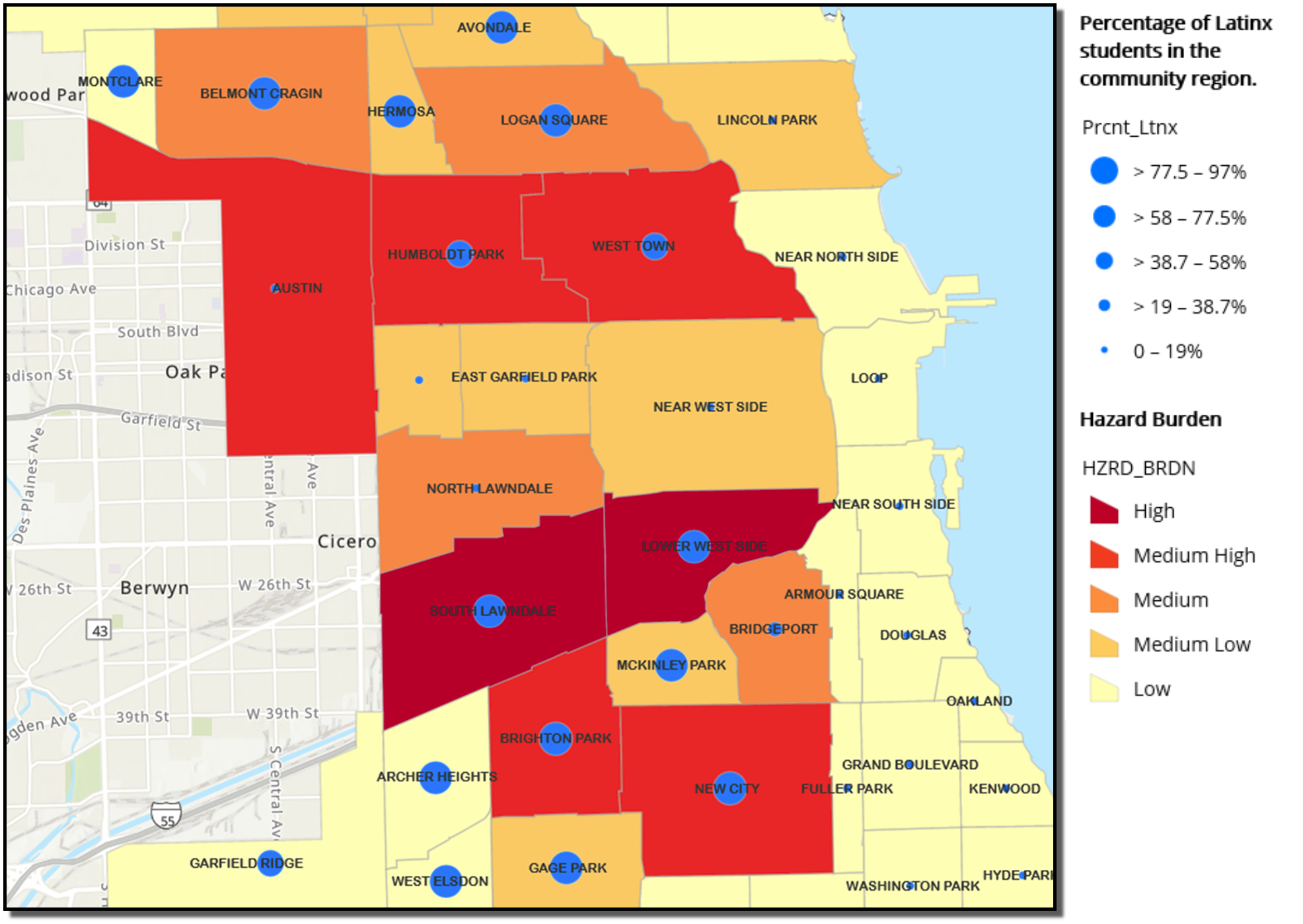}
    \caption{CPB scores at a CA level for the central section of Chicago. Regions are shaded according to their hazard burden scores. Blue circles highlight the percentage of Latinx students.}
    \vspace{-0.8cm}
    \label{fig:spatial}
\end{wrapfigure}

As seen in Fig.~\ref{fig:spatial}, the burden classification from this source is not randomly distributed within the City of Chicago, and a few community areas are allotted the highest-burden score. In addition, the industrial road burden score is concentrated in communities with a predominantly Latinx student population (i.e., more than 58\% in all the High burden CPB categories). The finding underlines the EJ implications since the LatinX student population and the percent of the overall Latinx population in each community area are highly correlated.

\begin{figure}[h]
    \centering
    \vspace{-0.5cm}
    \includegraphics[width=0.45\textwidth]{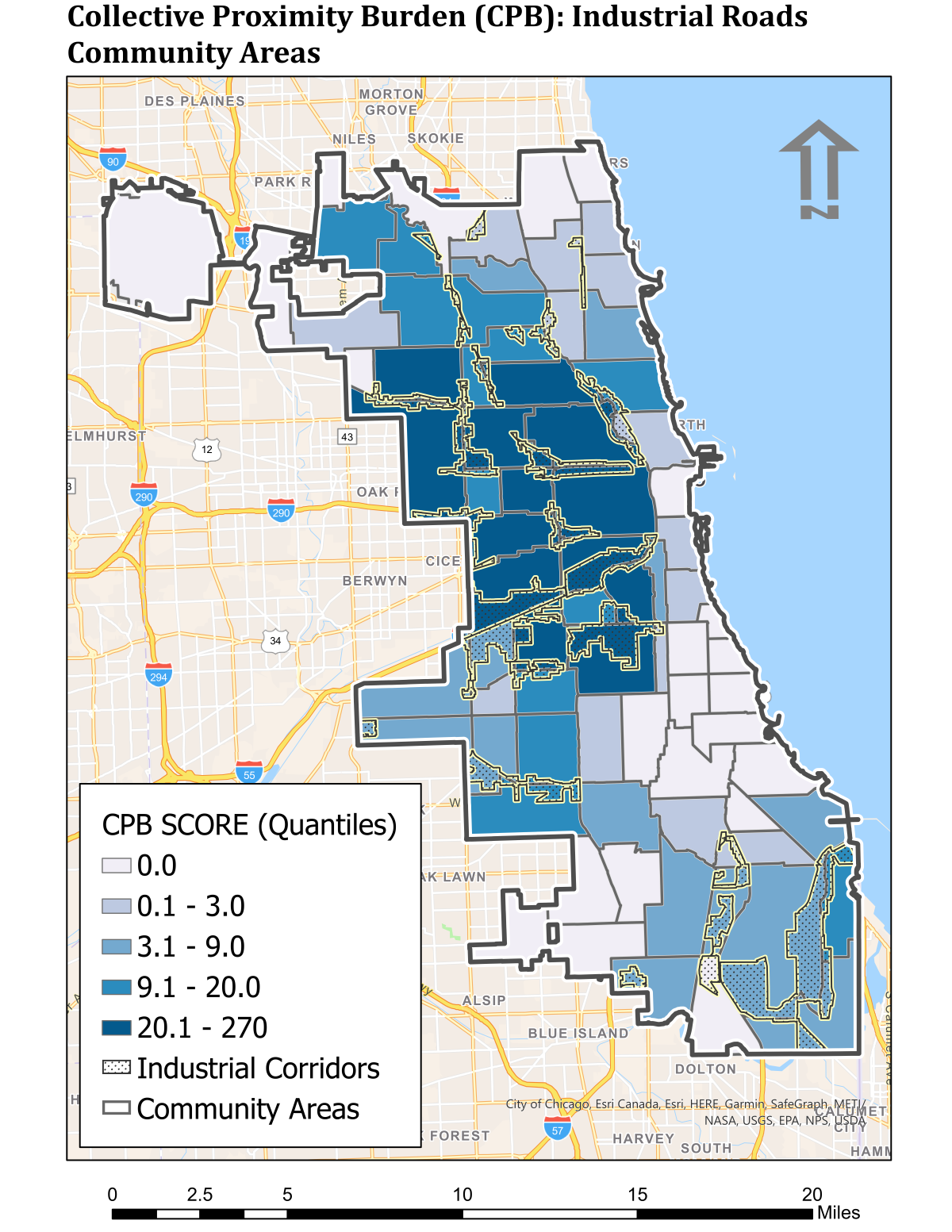}
    \includegraphics[width=0.45\textwidth]{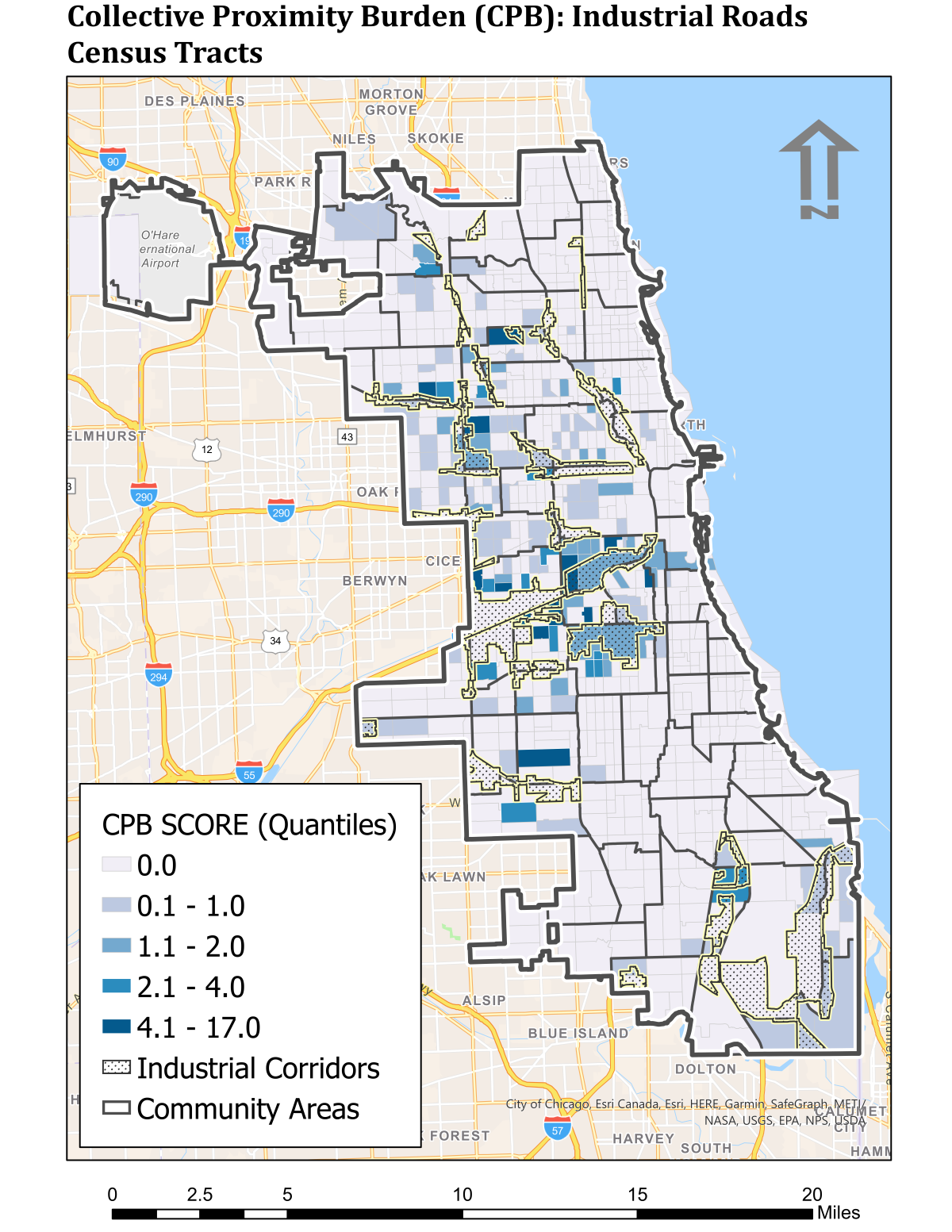}
    \caption{Left: Industrial roads CPB distribution at a CA scale. Right: Industrial roads CPB distribution at a CT scale.}
    \vspace{-0.75cm}
    \label{fig:spatial2}
\end{figure}

The community area (CA) scale of aggregation and the classification based on natural breaks were selected as the most ``sensible'' representations of the EJ issues during our interactions with local community groups and policymakers.  However, when spatial data are aggregated at a CA scale, the results are likely to depend on the selected scale and the configuration of the areal units adopted to represent the burden scores. In our studies of EJ issues in Chicago the modifiable areal unit problem (MAUP)~\cite{wong2004modifiable} was identified, however, the major conclusion is not altered since the burden distribution at a census tract (CT) level across the city remains unequal and much more evident. In addition, the CTs with the high burden classification contain a predominantly Latinx student population (results not shown).  For comparison purposes, a quantile classification was implemented for depicting the industrial roads burden distribution at a CA and CT scale. Fig.~\ref{fig:spatial2} corroborates the above findings and further justify the use of natural breaks as a much more conservative classification method. The association of the CPB scores and industrial corridors becomes evident as well since the majority of the high-level scores are concentrated near these corridors.

The high-burden CTs provide a practical resolution with a better delineation of the critical areas. This resolution is helpful for designing and implementing environmental monitoring networks, identifying restraint areas for new TRI level industrial developments and permits, and identifying priority areas for rezoning the industrial corridors.

\vspace{-0.3cm}
\section{Conclusion}
\vspace{-0.3cm}

Through the conscious planning and support of the City of Chicago, the number of facilities in the surrounding industrial corridors is increasing, along with levels of pollution and health effects like increasing rates of cancer, asthma and respiratory disease. This lead to a growing dispute between the communities, developers and the City Council over rezoning decisions that will directly impact the number of industries (and jobs) in the region as well as pollution levels.
A major innovation of this study is the use of public school children to quantify the proximity burden in communities. 
The case study presented here indicates that certain community areas with a predominantly Latinx population are likely to bear a disproportionate burden of environmental pollution caused by industrial roads.
The adopted community-based participatory design approach for creating visualizations of EJ issues generated dashboards and findings that assist community groups and policymakers to gain operational insights and promote data-driven decisions to resolve these issues.
In addition, these dashboards provide the means to identify high priority areas for monitoring and rezoning efforts.

In future work, we plan to leverage the engagement with the community to develop alternate scenarios based on land use change~\cite{Doraiswamy:2018:IVE:3183713.3193559}, and effectively evaluate the impact of these changes on different features, such as transportation and population density~\cite{liu2020gmel,peregrino2021transportation}.
We based our calculation of hazard indicators on temporal homogeneity assumption and treated all hazards as the same, while in real-world scenarios, different hazards can be substantively different across a profile of relevant factors. In our future work, we will develop models that assign different weights to various type of hazards to create a more realistic representation. 
We also plan to incorporate visualizations at the census tract level to provide a better granularity of high-priority areas, and consider other minorities other than Latinx.
Given the wealth of urban data currently available, we also plan to reproduce our work in other US cities.

\vspace{-0.3cm}
\bibliographystyle{splncs04}
\bibliography{references}

\begin{thebibliography}{10}
\providecommand{\url}[1]{\texttt{#1}}
\providecommand{\urlprefix}{URL }
\providecommand{\doi}[1]{https://doi.org/#1}

\bibitem{brender2011residential}
Brender, J.D., Maantay, J.A., Chakraborty, J.: Residential proximity to
  environmental hazards and adverse health outcomes. American journal of public
  health  \textbf{101}(S1),  S37--S52 (2011)

\bibitem{Doraiswamy:2018:IVE:3183713.3193559}
Doraiswamy, H., Tzirita~Zacharatou, E., Miranda, F., Lage, M., Ailamaki, A.,
  Silva, C.T., Freire, J.: Interactive visual exploration of spatio-temporal
  urban data sets using {Urbane}. In: Proc. 2018 Int. Conf. on Management of
  Data. SIGMOD '18

\bibitem{MCVDa}
Flax-Hatch, J., Sambanis, A., Cailas, M.: {MCVD: Environmental Justice and
  Neighborhood Schools in Chicago, Illinois. Part 1}  (2021)

\bibitem{MCVDb}
Flax-Hatch, J., Sambanis, A., Cailas, M.: {MCVD: Environmental Justice and
  Neighborhood Schools in Chicago, Illinois. Part 2}  (2021)

\bibitem{ChicagoTribune}
Husain, N.: {McKinley Park residents want asphalt plant shut down -- Chicago
  Tribune}. Retrieved from:
  \url{https://www.chicagotribune.com/news/environment/ct-environmental-justice-mckinley-park-asphalt-plant-20210528-bc352axgnzbqtlxf4yw6tj64nu-story.html}
  (2021)

\bibitem{liu2020gmel}
Liu, Z., Miranda, F., Xiong, W., Yang, J., Wang, Q., Silva, C.T.: Learning
  geo-contextual embeddings for commuting flow prediction. In: Proc.
  Thirty-Fourth AAAI Conference on Artificial Intelligence (2020)

\bibitem{doi:10.1080/23251042.2016.1220849}
Mah, A.: Environmental justice in the age of big data: challenging toxic blind
  spots of voice, speed, and expertise. Environmental Sociology  \textbf{3}(2),
   122--133 (2017)

\bibitem{peregrino2021transportation}
Peregrino, A.A., Pradhan, S., Liu, Z., Ferreira, N., Miranda, F.:
  Transportation scenario planning with graph neural networks (2021)

\bibitem{us2015human}
{US EPA}: Human health risk assessment: Strategic research action plan
  2016-2019. EPA 601/K-15/002  (2015)

\bibitem{epajustice}
{US EPA}: {Environmental Justice}. Retrieved from:
  \url{https://www.epa.gov/environmentaljustice} (2021)

\bibitem{wong2004modifiable}
Wong, D.W.: The modifiable areal unit problem {(MAUP)}. In: WorldMinds:
  Geographical perspectives on 100 problems, pp. 571--575. Springer (2004)

\end{thebibliography}
\end{document}